\documentclass[preprint,unsortedaddresss,howpacs,preprintnumbers,amsmath,amssymb]{revtex4}
\usepackage[colorlinks, citecolor=blue, urlcolor=blue, linkcolor=blue]{hyperref}
\usepackage{booktabs}
\usepackage{mathrsfs}
\usepackage{epsfig}
\usepackage{graphicx}
\usepackage{dcolumn}
\usepackage{bm}
\usepackage{amsmath}
\usepackage{slashed}
\usepackage{epstopdf}
\usepackage{inputenc}
\usepackage{multirow}
\usepackage{amsmath}		
\usepackage{amssymb}		
\usepackage{color}
\def\beqa{\begin{Eqnarray}}
\def\eeqa{\end{Eqnarray}}
\def\beq{\begin{equation}}
\def\eeq{\end{equation}}

\begin{document}

\title{Probing GeV-scale MSSM neutralino dark matter in collider and direct detection experiments}

\author{ Guang Hua Duan$^{1,2,3}$}
\author{ Wenyu Wang$^{4}$}
\author{ Lei Wu$^{1}$}
\author{ Jin Min Yang$^{2,3,5}$}
\author{ Jun Zhao$^{4,2}$
\vspace*{.5cm}}

\affiliation{
$^1$ Department of Physics and Institute of Theoretical Physics, Nanjing Normal University, Nanjing, Jiangsu 210023, China\\
$^2$ CAS Key Laboratory of Theoretical Physics, Institute of Theoretical Physics, Chinese Academy of Sciences, Beijing 100190, China\\
$^3$ School of Physical Sciences, University of Chinese Academy of Sciences, Beijing 100049, China\\
$^4$ Institute of Theoretical Physics, College of Applied Science, Beijing University of Technology, Beijing 100124, China\\
$^5$ Department of Physics, Tohoku University, Sendai 980-8578, Japan
}%

\begin{abstract}
Given the recent constraints from the dark matter (DM) direct detections, we examine a light GeV-scale (2-30 GeV)
neutralino DM in the alignment limit of the Minimal Supersymmetric Standard Model (MSSM).
In this limit without decoupling, the heavy CP-even scalar $H$ plays the role of the Standard Model (SM)
Higgs boson while the other scalar $h$ can be rather light so that the DM can annihilate through
the $h$ resonance or into a pair of $h$ to achieve the observed relic density.
With the current collider and cosmological constraints, we find that such a light neutralino DM above
6 GeV can be excluded by the XENON-1T (2017) limits while the survivied parameter space below 6 GeV
can be fully tested by the future germanium-based light dark matter detections (such as CDEX),
by the Higgs coupling precison measurements or by the production process $e^+e^- \to hA$ at an
electron-positron collider (Higgs factory).
\end{abstract}
\maketitle

\section{Introduction}\label{sec:intro}

The existence of cold dark matter (CDM) has been confirmed by many astrophysical experiments.
The CDM provides a natural way to account for many properties of galaxies and its predictions are in good
agreement with data on large scales. The nature of CDM, however, has remained elusive and been extensively
studied in particle physics. Among various hypothesis for CDM, the most compelling one is
the Weakly Interacting Massive Particle (WIMP) in the mass range 10-1000 GeV.

So far, the WIMP dark matter has undergone stringent experimental scrutiny. The direct DM detections
attempt to measure the nuclear recoil imparted by the scattering of DM. The recent strong limits on
nucleon-WIMP scattering have already excluded (at 90\% CL) a spin-dependent cross section
above $4.1 \times 10^{-41}$ for a WIMP mass of 40 GeV~\cite{pandaX-II} and a spin-independent
cross section above $7.7 \times 10^{-47}$ for a WIMP mass of 35 GeV~\cite{xenon-1t}, approaching
the neutrino floor. Besides, the null results from indirect DM detections via gamma rays, positrons
and neutrinos as well as collider searches for mono-X signatures also put stringent constraints
on various WIMP candidates~\cite{dm-review}.

The MSSM is one of the most popular extensions of the SM, in which the lightest neutralino can be a natural
WIMP dark matter if $R$-parity is conserved. With the LHC data, the precision electroweak data and flavour
measurements as well as the dark matter detection limits, the typical WIMP mass range has been largely
excluded, albeit some blind spots in direct detections are still remained due to some accidental cancellation
in the couplings of Higgs/$Z$ boson with the neutralino DM. A lower mass limit of the bino-like DM
in the 19-parameter MSSM is found to be about 30 GeV when the lighter CP-even Higgs boson ($h$) is
SM-like~\cite{atlas,wu-1,dev-1}. On the other hand, due to accidental cancellation effects in the alignment
limit without decoupling \cite{align-1,align-2,align-3,align-4,dev-2},
the heavier CP-even scalar ($H$) can serve as the observed 125 GeV Higgs boson
while the other scalar $h$ can be very light. This provides a possibility that the bino-like DM below
30 GeV can saturate the DM relic density with the help of a light $h$~\cite{Profumo:2016zxo,Ren}.
Since current direct detection experiments have generally poor sensitivities to a light GeV-scale DM,
this scenario may be still viable and worth a thorough study.

In this work, we examine such a light bino-like DM in the MSSM with $H$ playing the role of the
SM-like Higgs boson.
In particular, we focus on the GeV-scale DM in a mass range of 2-20 GeV, in which DAMA~\cite{dama},
COGENT~\cite{cogent} and CRESST~\cite{cresst} experiments have reported some plausible signals
of WIMP interaction. We will first utilize the current LHC Run-2 data and DM direct detection results to examine
whether such a GeV-scale neutralino DM is still allowed in the MSSM. Then we will explore the prospect of probing
such a light DM by the projected germanium-based light dark matter detectors, by the Higgs coupling precision
measurements and by the production process $e^+e^- \to A(\to b\bar{b})h(\to b\bar{b})$ at
future $e^+e^-$ colliders.

The structure of this paper is organized as follows. In Section \ref{section2}, we will briefly describe
the alignment limit of the Higgs sector in the MSSM and its implications on the phenomenology of the
light neutralino DM. In Section \ref{section3}, we confront our light DM scenario with the current LHC
and DM experimental data, and investigate its related phenomenologies in future collider and direct
detection experiments. Finally, we draw our conclusions in Section \ref{section4}.

\section{MSSM Higgs alignment limit and implications on DM}\label{section2}
The MSSM has a minimal Higgs sector consisting of two Higgs doublets with opposite hypercharges \cite{MSSM}:
\begin{eqnarray}
H_u = \left(\begin{array}{c} H_u^+ \\ H_u^0 \end{array} \right), \qquad
H_d = \left(\begin{array}{c} H_d^{0*} \\ -H_d^- \end{array} \right) \, .
\end{eqnarray}
The tree-level Higgs potential is given by
\begin{eqnarray}
V & = & (|\mu|^2+m_{H_u}^2) |H_u|^2 +(|\mu|^2+m_{H_d}^2) |H_d|^2 - B \mu (\epsilon_{\alpha\beta} H_u^\alpha H_d^\beta
+ h.c.) \nonumber \\
& & +\frac{g^2+g'^2}{8} (|H_u|^2-|H_d|^2)^2 + \frac{g^2}{2}
|H_u^\dagger H_d|^2 \; ,
\end{eqnarray}
where $\mu$ is the higgsino mass parameter, $B$ is the soft SUSY-breaking bilinear Higgs term,
$m_{H_u}^2$ and $m_{H_d}^2$ are the Higgs mass parameters squared, $g$ and $g'$ are the SM $SU(2)$
and $U(1)$ gauge couplings. With spontaneous breaking of electroweak symmetry, the neutral components
of the two Higgs doublets both develop vacuum expectation values $v_{u,d}$, whose squared sum is
$v_u^2+v_d^2=v^2$ with $v\approx 246$ GeV and their ratio $\tan\beta=v_u/v_d$ is a free parameter.
A neutral Goldstone $G^0$ from the neutral components and a pair of charged
Goldstones $G^\pm$ from the charged components are eaten by gauge bosons $Z$ and $W^\pm$, respectively.
The remained degrees of freedom give two CP-even Higgs bosons $h$ and $H$ ($m_h<m_H$), a CP-odd Higgs
boson $A$ and two charged Higgs bosons $H^\pm$.

In order to show the alignment limit, we use the Higgs basis $(H_1,H_2)$ defined as \cite{higgs-basis}
\begin{align}
H_1 = \begin{pmatrix} H_1^+\\H_1^0 \end{pmatrix} \equiv \frac{v_1 \Phi_1 + v_2\Phi_2}{v}, \qquad H_2 = \begin{pmatrix} H_2^+\\H_2^0 \end{pmatrix} \equiv \frac{-v_2 \Phi_1 + v_1\Phi_2}{v}
\end{align}
with
\beq
(\Phi_1)^i=\epsilon_{ij}(H_d^*)^j\,,\qquad\quad (\Phi_2)^i=(H_u)^i\,,
\eeq
where $\epsilon_{12}=-\epsilon_{21}=1$ and $\epsilon_{11}=\epsilon_{22}=0$, and there is an implicit sum over
the repeated $SU(2)$ index $j=1,2$. It can be seen that $\langle H_1^0 \rangle = v/\sqrt{2}$ and
$\langle H_2^0\rangle = 0$.  The CP-even mass eigenstates are related with the neutral Higgs mass
eigenstates $\{\sqrt{2}\,{\rm Re}~H^0_1-v$\,,\,$\sqrt{2}\,{\rm Re}~H^0_2\}$ by
\beq \label{mixing}
\begin{pmatrix} H\\ h\end{pmatrix}=\begin{pmatrix} c_{\beta-\alpha} & \,\,\, -s_{\beta-\alpha} \\
s_{\beta-\alpha} & \,\,\,\phantom{-}c_{\beta-\alpha}\end{pmatrix}\,\begin{pmatrix} \sqrt{2}\,\,{\rm Re}~H_1^0-v \\
\sqrt{2}\,{\rm Re}~H_2^0
\end{pmatrix}\,,
\eeq
where $c_{\beta-\alpha} \equiv\cos(\beta - \alpha)$ and $s_{\beta-\alpha} \equiv\sin(\beta - \alpha)$ are defined in
terms of the mixing angle $\alpha$ that diagonalizes the CP-even Higgs squared-mass matrix when expressed in
the original basis of scalar fields, $\{\sqrt{2}\,{\rm Re}~\Phi_1^0-v_1\,,\,\sqrt{2}\,{\rm Re}~\Phi_2^0-v_2\}$.

In terms of the Higgs basis fields, we can rewrite the Higgs potential as
\begin{align}
V = \ldots + \frac{1}{2} Z_1 (H^\dagger_1 H_1)^2 + \ldots +[\frac{1}{2} Z_5 (H^\dagger_1 H_2)^2 + Z_6 (H^\dagger_1 H_1)(H^\dagger_1 H_2) + {\rm h.c.}]+ \ldots
\end{align}
At tree-level, the above quartic couplings $Z_1$, $Z_5$ and $Z_6$ are given by
\beq
Z_1=\tfrac{1}{4}(g^2+g^{\prime\,2}) c_{2\beta}^2\,,\qquad\quad
Z_5=\tfrac{1}{4}(g^2+g^{\prime\,2}) s_{2\beta}^2\,,\qquad\quad
Z_6=-\tfrac{1}{4}(g^2+g^{\prime\,2}) s_{2\beta}c_{2\beta}
\eeq
where $c_{2\beta}\equiv\cos 2\beta$ and $s_{2\beta}\equiv \sin 2\beta$. Then, we can compute the squared-mass matrix of the neutral CP-even Higgs bosons, with respect to the neutral Higgs states, $\{\sqrt{2}\,{\rm Re}~H^0_1-v$\,,\,$\sqrt{2}\,{\rm Re}~H^0_2\}$
\begin{eqnarray}
\mathcal{M}^2=\left( \begin{array}{cc}
Z_1 v^2& Z_6 v^2 \\
Z_6 v^2& M^2_A + Z_5 v^2
\end{array} \right).
\label{mass}
\end{eqnarray}
If $\sqrt{2}\,{\rm Re}H^0_1-v$ were a Higgs mass eigenstate, it would have the same couplings as the SM Higgs boson at tree level. In other words, to obtain a SM-like Higgs boson, one of the neutral Higgs mass eigenstates has to be close to $\sqrt{2}\,{\rm Re}H^0_1-v$. From Eq.~(\ref{mixing}) and ~(\ref{mass}), we can find that
\begin{enumerate}
\item
\textit{Decoupling limit}, $M^2_A+Z_5 v^2 \gg Z_1 v^2$. In this case $h$ is SM-like and $M_A\sim M_H \sim M_{H^\pm}\gg M_h$.
\item
\textit{Alignment limit without decoupling}, $|Z_6|\ll 1$. In this case $h$ is SM-like if $(M^2_A+Z_5 v^2) > Z_1 v^2$ ~\cite{yun-1} and $H$ is SM-like if $M^2_A+ Z_5 v^2 < Z_1 v^2$ ~\cite{yun-2}. The latter case necessarily corresponds to this alignment limit.
\end{enumerate}
It should be noted that the exact alignment without decoupling, $Z_6=0$, trivially occurs when $\beta=0$ or $\pi/2$ (corresponding to the vanishing of either $v_1$ or $v_2$). However, this will lead to a massless $b$ quark or $t$ quark, respectively, at tree-level. Therefore, the MSSM Higgs alignment $Z_6=0$ can only happen through an accidental cancellation of the tree-level terms with contributions arising at the one-loop level (or higher). In the limit $M_{Z,A} \ll M_S$, the leading one-loop correction to $Z_6$ is given by
\begin{align}
Z_6v^2 &= -s_{2\beta} \left\{M_Z^2c_{2\beta}  - \frac{3 m_t^4 }{4\pi^2v^2s_\beta^2} \left[\ln\left(\frac{M_S^2}{m_t^2}\right) +\frac{X_t(X_t+Y_t)}{2 M_S^2} - \frac{X_t^3Y_t}{12M_S^4} \right]\right\},
\label{Eq:Z6v2}
\end{align}
where $M_S\equiv \sqrt{m_{\tilde{t}_1}m_{\tilde{t}_2}}$, $X_t \equiv A_t - \mu/\tan\beta$ and $Y_t \equiv A_t + \mu\tan\beta$. In Ref.~\cite{align-3}, authors numerically solved Eq.~(\ref{Eq:Z6v2}) and found that a large value of $\mu$ is required to achieve the alignment limit without decoupling when $\tan\beta$ is small. Note that the radiative corrections can also affect the SM-like Higgs couplings, in particular the Yukawa couplings of the third generation fermions~\cite{align-1,align-2,wu-2,wu-3}. The precision measurements of them will further test the MSSM alignment limit.

Since the Higgs alignment is independent of $M_A^2$, $Z_1$ and $Z_5$, the lighter CP-even Higgs boson $h$ can be light if the heavy Higgs boson $H$ is interpreted as the SM-like Higgs boson. The appearance of light Higgs boson $h$ will enrich MSSM dark matter phenomenology. As known, a light bino-like neutralino dark matter might overclose universe. Several possible ways for it being a viable thermal relic consistent with the observed abundance have been found, such as a mixture of higgsino and bino in natural SUSY~\cite{wu-1,wu-4}. While in the alignment limits, the light Higgs boson $h$ can play the role of mediator. The light bino-like DM can annihilate through it into the SM particles. Besides, they can also directly decay into a pair of $h$. This pushes the boundaries of how light the cold thermal neutralino relic can be in the MSSM. In the following, we will examine this possibility and focus on the GeV scale neutralino DM.

\section{Phenomenological studies and results }\label{section3}
We scan the relevant parameters in the following ranges:
\begin{eqnarray}
2{\rm ~GeV} \leq M_1 \leq 30{\rm ~GeV}, \quad 2 {\rm ~TeV} \leq \mu \leq 10 {\rm~TeV}, \quad 1 \leq \tan\beta \leq 50 \nonumber \\
-3 {\rm ~TeV} \leq A_t=A_b \leq 3 {\rm ~TeV}, \quad 1 {\rm ~TeV} \leq M_{Q_3}=M_{U_3}=M_{D_3} \leq 5 {\rm ~TeV}.\label{parameter}
\end{eqnarray}
Here, the lower value of $ M_1$ is inspired by the Lee-Weinberg bound of WIMP dark matter \cite{lee}. It should be mentioned that the bino-like DM in 19-parameter pMSSM with $h$ being the SM-like Higgs boson has to be heavier than 30 GeV \cite{atlas}. However, this conclusion may not be valid in the alignment limit with $H$ being the 125 GeV Higgs boson, which motivates our upper value of $M_1$. In order to realize alignment limit at reasonably low $\tan \beta$ values that are experimentally allowed, one needs $\mu/M_S \geq O(2 \sim3)$ so that we require $\mu \geq 2$ TeV~\cite{align-3}. Besides, we demand the heavy stops and/or large Higgs-stop trilinear soft-breaking coupling to achieve the correct Higgs mass.
It is noted that the vacuum stability may give a very strong bound on the large value of $\mu/M_S$, which is considered in our study by using the approximate formulae in Ref.~\cite{hollik}. Given the LEP and LHC bounds on the first two generation squarks, sleptons, gluino and electroweakinos, we set $M_3= 3$ TeV, $M_2=3$ TeV, $A_{\tau} = -100$ GeV, $M_{Q_{1,2}} =M_{U_{1,2}}=M_{D_{1,2}}= 3$ TeV, $M_{L_{1,2}}=M_{E_{1,2}}=1$ TeV and $M_{L_3}=M_{E_3}=1.5$ TeV for simplicity. We also take $m_{H^\pm} =155{\rm ~GeV}$ and the top quark mass $M_t=173.2{\rm ~GeV}$ in our calculations. We evaluate our parameter space by the usual random scan of Eq.~\ref{parameter}. The surviving possibility of our samples is about 1 out of 6 million. The advanced numerical techniques, such as MultiNest sampling, may improve our scan efficiency.

In our scan we consider the following experimental constraints:
\begin{itemize}
\item[(1)] We use \textsf{FeynHiggs-2.13.0}~\cite{feynhiggs} to calculate the Higgs mass, including various two-loop corrections and NNLL resummation contributions. Since we use $m_{H^\pm}$ as an input parameter, we set the FeynHiggs flag `tlCplxApprox=1'. We require the heavy CP-even Higgs boson $H$ to be the SM-like Higgs boson with its mass in the range of $122 < m_H< 129$ {\rm ~GeV}.

\item[(2)] We impose the exclusion limits (at the 95\% confidence level) from LEP, Tevatron and LHC in Higgs searches with \textsf{HiggsBounds-4.3.1}~\cite{higgsbounds}. For low $M_A$ values, the $H/A \to \tau^+\tau^-$ search results from the combined 7 and 8 TeV run are still more sensitive than the current 13 TeV results. We also perform the Higgs data fit by calculating $\chi^2$ of the Higgs couplings with \textsf{HiggsSignals-1.4.0} \cite{higgssignals} and require $\chi^2 \lesssim 112.7273$ (corresponding to 95\% C.L.) for the number of observables $N_{obs}=89$. We choose the SLHA input choice of HiggsBounds/HiggsSignals, where the effective Higgs couplings are only used to calculate the Higgs production cross section ratios.

\item[(3)] We calculate the thermal relic density of the lightest neutralino (as the dark matter candidate) by \textsf{MicrOMEGAs 4.3.5} \cite{micromegas} and require its value within $3\sigma$ range of the Planck observed value, $\Omega_{\rm DM}h^2=0.1186\pm0.03651$ \cite{planck}.
\end{itemize}

\begin{figure}[h]
\centering
\includegraphics[width=15cm,height=8cm]{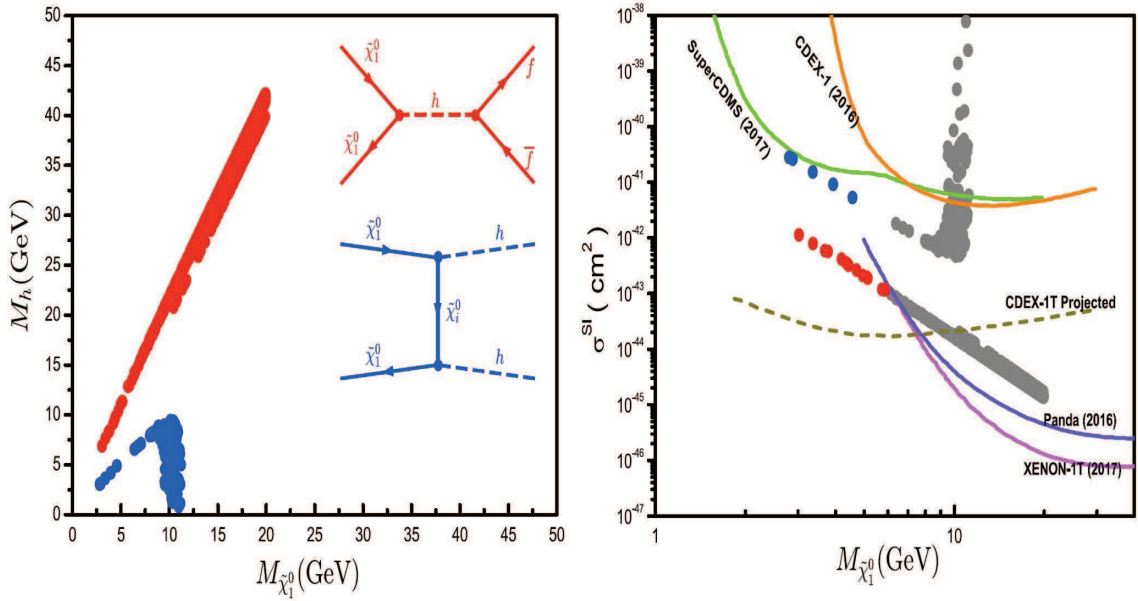}
\vspace{-.3cm}
\caption{Left panel shows the samples satisfying constraints (1)-(3) on the plane of $m_{\tilde\chi_1^0} $
versus $m_ h$. There are two ways to achieve the correct DM relic density: one is that DMs annihilate
into the SM particle (red dots) through $s$-channel mediated by a light $h$ boson; the other is that
DMs directly annihilate into a pair of $h$ bosons (blue dots). Right panel shows spin-independent
neutralino LSP-nucleon scattering cross section, where the samples are same as in the left panel
(those samples excluded by current direct detections are plotted in gray).
The observed 90\% CL upper limits from CDEX-1(2016)~\cite{cdex2016},
SuperCDMS(2017)~\cite{SuperCDMS-2017}, PandaX(2016)~\cite{pandax2016} and
XENON-1T(2017)~\cite{xenon-1t}, and the future projected CDEX-1T~\cite{cdex} limits
are shown.}
\label{dm}
\end{figure}
In Fig.~\ref{dm} we display samples satisfying constraints (1)-(3) on the plane of $m_{\tilde\chi_1^0} $
versus $m_h$. We can see that there are two ways to satisfy the requirement of observed DM relic
density: one is that DMs annihilate into the SM particle through $s$-channel mediated by a light $h$
boson  (red dots), which requires that $m_h$ is about twice of $m_{\tilde\chi^0_1}$; the other is that
DMs directly annihilate into a pair of $h$ bosons through the $t$-channel (blue dots). The DM mass
$m_{\tilde\chi^0_1}$ can be as low as about 2 GeV.
As we can see, the surviving samples in the left panel have been tightly constrained by current
limits from DM direct detection experiments. The neutralino DM with
$6~{\rm GeV} \lesssim m_{\tilde{\chi}_1^0} \lesssim 30 ~{\rm GeV}$ is excluded by the XENON-1T (2017) limits.
Due to advantages of low energy threshold and good energy resolution, the germanium detectors are
the effective ways to probe the light WIMP. So, it can be seen that the projected CDEX-1T\cite{cdex}
will test all our samples in the future.

\begin{figure}[h]
\centering
\includegraphics[width=15cm,height=8cm]{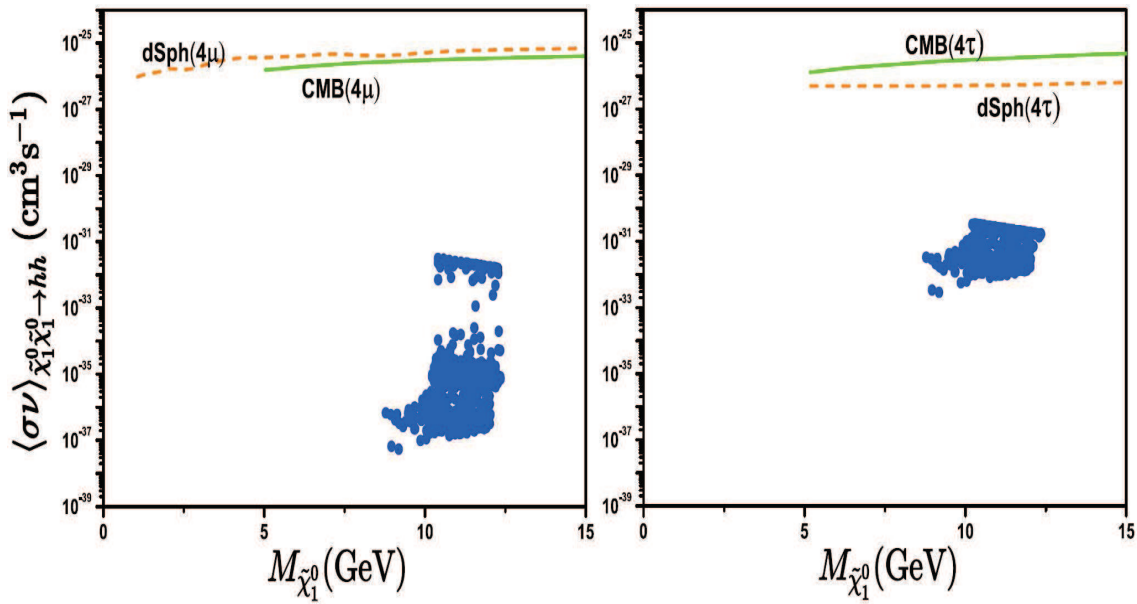}
\vspace{-.5cm}
\caption{Same as Fig.~\ref{dm}, but showing 95\% C.L. upper limits from Fermi-LAT (orange dashed lines) and Planck data (green solid lines) on $\langle \sigma v \rangle_{\tilde{\chi}^0_1 \tilde{\chi}^0_1 \to hh}$ for $4\mu$ and $4\tau$ final states~\cite{stefano}.}
\label{ID}
\end{figure}

In Fig.\ref{ID}, we show 95\% C.L. upper limits from Fermi-LAT and Planck data on the thermally-averaged $t$-channel cross-section $\langle \sigma v \rangle_{\tilde{\chi}^0_1 \tilde{\chi}^0_1 \to hh}$ for $4\mu$ and $4\tau$ final states~\cite{stefano}. We can see that our samples are far below those bounds because of suppressions of Higgs branching ratios. It should be mentioned that there is now no corresponding limit for $b\bar{b}$ final states when $m_{\tilde{\chi}^0_1} < 15$ GeV~\cite{stefano}. Moreover, the $s$-channel cross section $\langle \sigma v \rangle_{\tilde{\chi}^0_1 \tilde{\chi}^0_1 \to h \to 2\mu,2\tau,2b}$ can also evade the indirect detection bounds \cite{Magic and Fermi} due to $p$-wave suppression.

\begin{figure}[h]
\centering
\includegraphics[width=15cm,height=15cm]{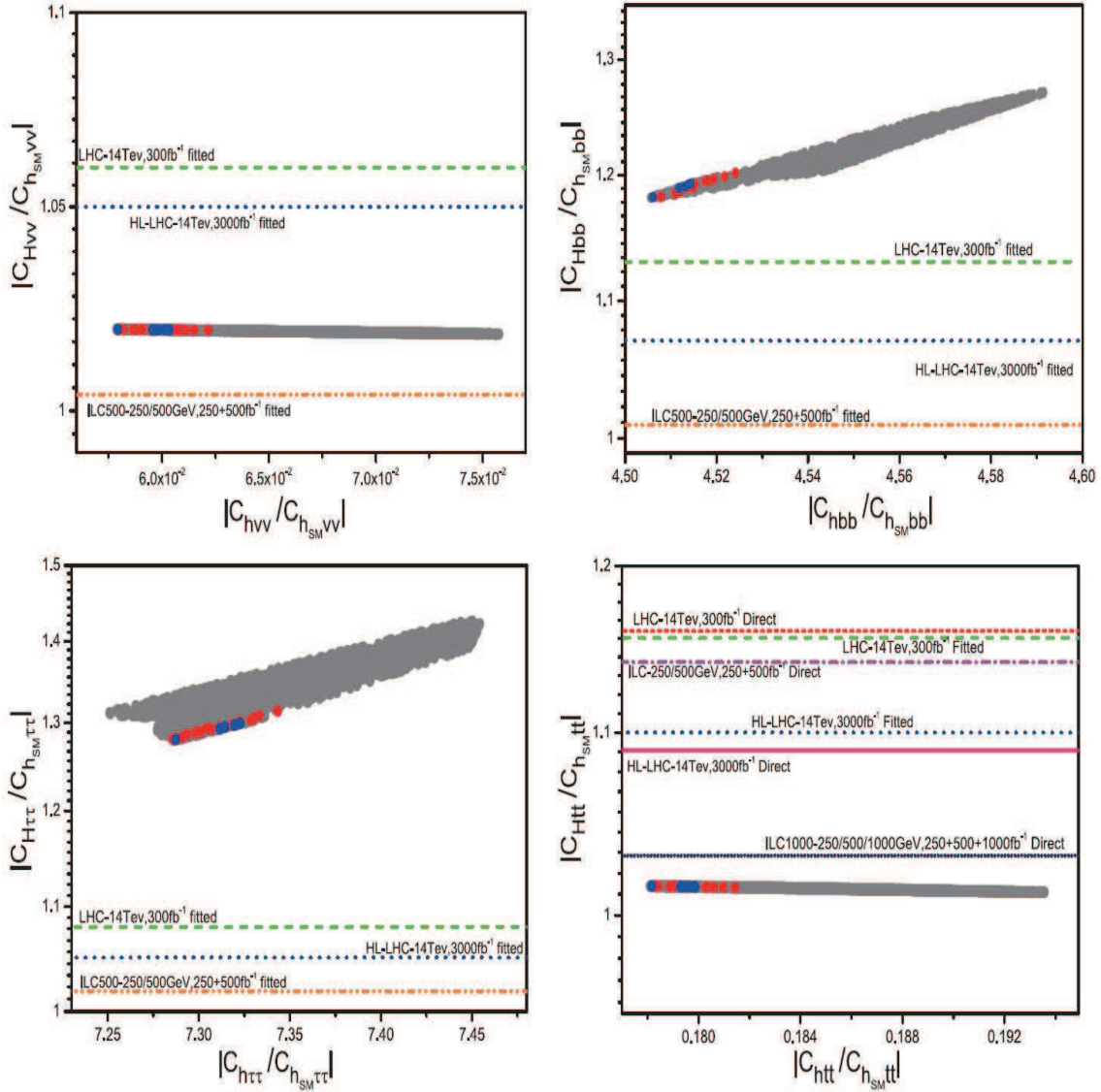}
\vspace{-.5cm}
\caption{Same as Fig.~\ref{dm}, but showing the Higgs couplings normalized to the SM values.
The expected LHC-14, HL-LHC, ILC sensitivities to the SM Higgs couplings ~\cite{ilc}
are also plotted (the region above each line is the corresponding observable region).}
\label{couplings}
\end{figure}

In Fig.\ref{couplings} we present the Higgs couplings normalized to the SM values.
It can be seen that the coupling $HVV$ ($V=Z,W$) is very close to the SM prediction
while the coupling $hVV$ is highly suppressed.
This verifies our scan results that $H$ is the SM-like Higgs boson.
For the heavy Higgs $H$,  the down-type Yukawa couplings $Hb\bar{b}$ and $H\tau^+\tau^-$ are
enhanced sizably, which can be tested in the future HL-LHC and ILC. On the other hand, the
up-type Yukawa coupling $Ht\bar{t}$ has a small deviation from the SM value and is below
the ILC sensitivity.

\begin{figure}[h]
\centering
\includegraphics[width=15cm,height=7cm]{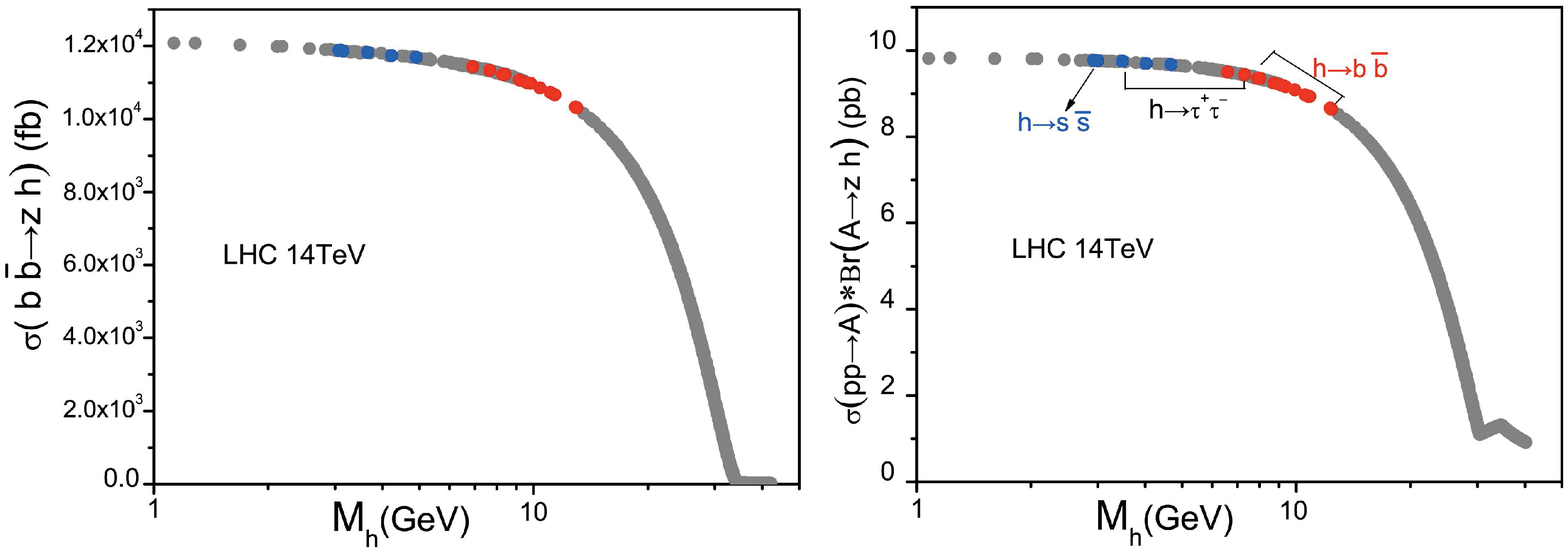}
\vspace{-.5cm}
\caption{Same as Fig.\ref{dm}, but showing the cross sections of the processes $b\bar{b} \to Zh$
and $gg+b\bar{b} \to A \to Zh$ at the 14 TeV LHC.}
\label{lhc}
\end{figure}
Since the light Higgs boson $h$ always accompanies the light DM as shown in Fig.~\ref{dm},
we calculate the production processes of $h$ at the LHC. In Fig.~\ref{lhc}, we show the cross
sections of $b\bar{b} \to Zh$ and $gg+b\bar{b} \to A \to Zh$ at the 14 TeV LHC, which for our
surviving samples can reach about 10-12 pb and 8-10 pb, respectively. The NLO QCD corrections
to the process $gg+b\bar{b} \to A$ are included by using $\textsf{SusHi-1.6.1}$~\cite{sushi}.
Through Monte Carlo simulations for the signals and backgrounds, we find these two
processes unobservable at the LHC because the decay products of $h$ is soft and
the mass splitting between the CP-odd Higgs boson $A$ and its final states $Zh$ is small.
Besides, the light Higgs boson $h$ can also be singly produced through gluon fusion
or $b\bar{b}$ annihilation, which, however, also suffers from the huge SM backgrounds
and is found unobservable at the LHC.

Next, we explore the potential of testing our light DM and light Higgs boson scenario
at a 240 GeV electron-positron collider (Higgs factory).
In Ref.~\cite{monoz,monophoton}, the authors studied mono-$Z$ and mono-photon signatures
in effective operator framework at an $e^+e^-$ collider. However, such production rates are
very small in our case since $\tilde{\chi}^0_1$ is very bino-like and its coupling with $Z$
boson is negligible weak.
Note that from the Higgs-gauge boson interactions in Eq.~(\ref{interaction}) we can see that
the coupling $|C_{hAZ}|$ is sizeable because of $\beta-\alpha \to \pi$ in the alignment limit
without decoupling:
\begin{eqnarray*}
    {\cal L} &\ni& -\frac{g_Z}{2}[ \cos (\alpha-\beta)(h\overleftrightarrow {\partial}^\mu A)+\sin (\alpha-\beta)(H\overleftrightarrow {\partial}^\mu A)]Z_\mu \\
              &&+\frac{g_Z}{2}M_Z[ \cos (\alpha-\beta)H+\sin (\alpha-\beta)h]Z_\mu Z^\mu.
    \label{interaction}
\end{eqnarray*}
Therefore, we investigate the observability of the process
$e^+e^- \to A(\to b\bar{b})h( \to b\bar{b})$ at a 240 GeV Higgs factory.
We generate parton level events using \textsf{MadGraph5\_aMC@NLO} \cite{Madgraph}.
The events are then passed to \textsf{Pythia} \cite{pythia} for showering and hadronization.
The detector simulation is implemented with \textsf{Delphes}\cite{delphes}, where we use the
CEPC detector card \cite{cepc-card}. The $b$-tagging efficiency is 80\%. The main SM backgrounds
includes $e^+e^- \to Z(\to b\bar{b})H(\to b\bar{b})$, $3j$ and $4j$.

\begin{figure}[h]
  \centering
  \includegraphics[width=6.5in]{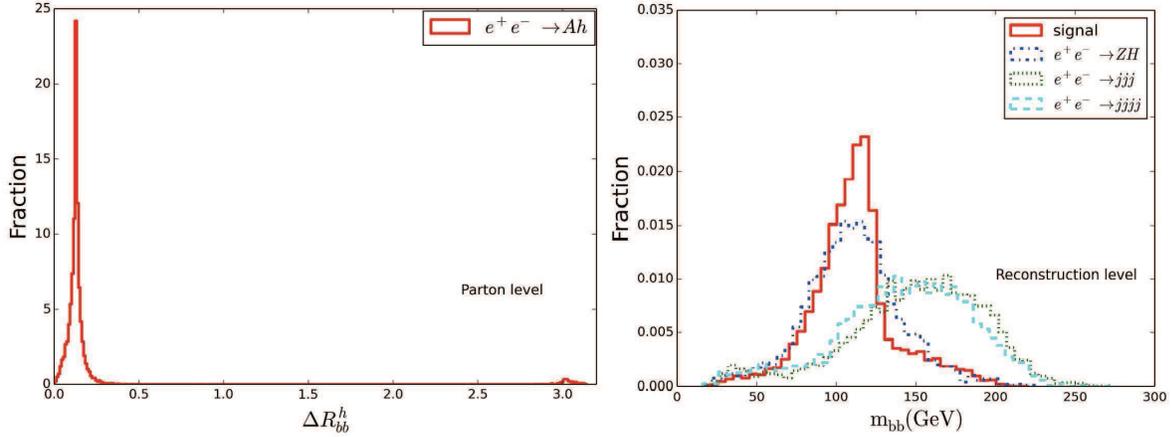}
\vspace{-1.0cm}
\caption{The distributions of $\Delta R_{bb}$ (parton level) and two $b$-jets
invariant mass $m_{bb}$ (reconstruction level). The signal benchmark point
is $(m_{A},m_{h})=(124.82,11.29)$ GeV.}
\label{fig:distribution}
\end{figure}
In Fig.~\ref{fig:distribution} we present the distributions of $\Delta R_{bb}$ (parton level)
and two $b$-jets invariant mass $m_{bb}$ (reconstruction level) at an $e^+e^-$ collider
with $\sqrt{s}=240$ GeV. We can see that the two $b$-jets from the decay of $h$ for most
of signal events are very collinear. Besides, we find that our signal and the
background $e^+e^- \to ZH$ have a peak around $m_{bb}=125$ GeV due to $m_A \simeq m_H$,
while the backgrounds $3j$ and $4j$ have a larger $m_{bb}$.

The detailed event selections are the followings:
\begin{itemize}
  \item We require exactly three jets with $p_{T} ^j>25$ GeV, $|\eta_j|< 3.5$ in the final states.
  Two of three jets are $b$-jets. The third jet is required to have $p_T>45$GeV.
  \item  We require the invariant mass $m_{bb}$ to be within 95-130 GeV.
\end{itemize}

\begin{table}[ht!]
\caption{A cut flow analysis for the cross sections (fb) of the backgrounds and signal at
the 240 GeV $e^+e^-$ collider with an integrated luminosity ${\cal L}=100 fb^{-1}$.
The benchmark point is $m_A=124.82$ GeV and $m_h=11.29$ GeV.}
\footnotesize\begin{tabular}{|c|c|c|}
  \hline
  cut & ~~$p^j_T>45$ GeV~~& ~~$95<m_{bb}<130$ GeV~~  \\
    \hline
  $ZH$ &9.08&4.43\\
  \hline
  $jjj$&49.50&14.05\\
  \hline
 ~~~~$e^+e^-\rightarrow h A$~~~~& 6.43& 4.12 \\
  \hline
  S/(S+B)(\%)&9.88&18.23\\
  \hline
  $S/\sqrt{S+B}$&7.96&8.67\\
  \hline
  \end{tabular}
\label{tab:cutflow}
\end{table}

In Table~\ref{tab:cutflow} we present a cut flow of the cross sections for the signal and backgrounds
at a 240 GeV $e^+e^-$ collider with an integrated luminosity ${\cal L}=100 fb^{-1}$. We can see that
the background $e^+e^- \to 3j$ is still about one order larger than the signal after the $p_T^j$ cut.
The requirement of $m_{bb}$ around $m_A$ can significantly suppress this background. The resulting
statistical significance $S/\sqrt{S+B}$ of the signal can reach $8.67\sigma$ with $S/B=18.23\%$.

\section{Conclusions}\label{section4}
In this work, we examined a light GeV-scale neutralino dark matter (2-30 GeV) in the MSSM.
Such a light WIMP DM can be realized in the Higgs alignment limit without decoupling,
which can pairly annihilate into the Standard Model particles through a light CP even Higgs boson
or into a pair of light CP even Higgs bosons to provide the correct relic density.
With the collider and cosmological constraints, we found that the DM with
$6 ~{\rm GeV} \lesssim m_{\tilde{\chi}_1^0} \lesssim 30 ~{\rm GeV}$ has been excluded by
the XENON-1T (2017) limit. By analyzing the surviving parameter space with
$2 ~{\rm GeV} \lesssim m_{\tilde{\chi}_1^0} \lesssim 6~{\rm GeV}$, we found that such
a light GeV-scale neutralino dark matter can be tested by the future
germanium-based light dark matter detectors (such as CDEX), by the Higgs coupling precision
measurements or by searching for the light Higgs boson $h$ through the process $e^+e^- \to hA$
at a Higgs factory.

\section*{Acknowledgement}
G. H. Duan was supported by a visitor program of Nanjing Normal University, during which this work was
finished. J. Zhao thanks Tim Stefaniak and Yang Zhang for technical supports and helpful discussions.
This work was supported by the National Natural Science Foundation of China (NNSFC)
under grant Nos. 11705093,  11305049, 11675242 and 11375001, by the CAS Center for Excellence in
Particle Physics (CCEPP), by the CAS Key Research Program of Frontier Sciences and
by a Key R\&D Program of Ministry of Science and Technology of China
under number 2017YFA0402200-04, by startup funding for Jiangsu Distinguished Professor.

\end{document}